\documentclass[prd,twocolumn,showpacs,preprintnumbers,floats,floatfix,apsrev,amsmath,amsfonts,nofootinbib,superscriptaddress]{revtex4-1}
\usepackage{color}
\usepackage{graphicx}
\usepackage{dcolumn}
\usepackage{natbib}
\usepackage{hyperref}
\usepackage{hypernat}
\usepackage{soul}
\usepackage{amssymb}

\newtheorem{thm}{Theorem}

\newcommand{\dd}{{\rm d}}
\newcommand{\be}{\begin{equation}}
\newcommand{\en}{\end{equation}}
\newcommand{\bea}{\begin{eqnarray}}
\newcommand{\ena}{\end{eqnarray}}
\newcommand{\bean}{\begin{eqnarray*}}
\newcommand{\enan}{\end{eqnarray*}}
\newcommand{\p}{\partial}
\newcommand{\riso}{\bar{{\rm r } } }
\newcommand{\yiso}{\bar{{\rm y } } }
\newcommand{\aiso}{\bar{{\rm a } } }
\newcommand{\biso}{\bar{{\rm b } } }

\usepackage{soul}

\begin{document}
\title{Entropy bound in Einstein-Born-Infeld black holes}
\author{F.~T.~Falciano}
\email{ftovar@cbpf.br} 
\affiliation{CBPF - Brazilian Center for Research in Physics, Xavier Sigaud st. 150, zip 22290-180, Rio de Janeiro, RJ, Brazil.}
\affiliation{PPGCosmo, CCE - Federal University of Esp\'\i rito Santo, zip 29075-910, Vit\'oria, ES, Brazil.}
\author{M.~L.~Pe\~{n}afiel}\email{mpenafiel@cbpf.br}
\affiliation{CBPF - Brazilian Center for Research in Physics, Xavier Sigaud st. 150,	zip 22290-180, Rio de Janeiro, RJ, Brazil.}
	\author{J.~C.~Fabris}\email{julio.fabris@cosmo-ufes.org}
\affiliation{PPGCosmo, CCE - Federal University of Esp\'\i rito Santo, zip 29075-910, Vit\'oria, ES, Brazil.}
\affiliation{N\'ucleo Cosmo-ufes \& Departamento de F\'\i sica - Universidade Federal do Esp\'\i rito Santo, zip 29075-910, Vit\'oria, ES, Brazil}
\affiliation{National  Research  Nuclear  University  MEPhI, Kashirskoe  sh. 31,  Moscow  115409,  Russia}

\begin{abstract}
We study the validity of Bekenstein's entropy bound for a charged black hole in the context of nonlinear electrodynamics. Bekenstein's inequalities are commonly understood as universal relations between the entropy, the charge, the momentum, and the energy of a physical system but independent of its dynamics. In particular, we consider the Born-Infeld electrodynamics coupled to gravity as described by General Relativity. Following the steps that lead to these inequalities, we study the absorption of a charged test particle by the black hole and verify that the entropy bound is violated. We find a modified upper bound for the entropy that depends on the maximum field parameter of the Born-Infeld theory.
\end{abstract}

\maketitle

\section{Introduction}

Black holes have been extensively studied in the literature and their general properties outlined. More recently, direct and indirect observations increased even more the interest in these astrophysical objects. One of the main ideas from black hole mechanics is to associate the horizon area $A$ with the black hole entropy $S$~\cite{Bekenstein1972, Wald1984} 
\[
S = \frac{k_B c^3}{\hbar G} \frac{A}{4} 
\quad ,
\]
where $k_B$ is the Boltzmann constant and $c$, $G$ and $\hbar$ are respectively the speed of light, the gravitational and Planck constants. The identification of the black hole area with its entropy allows for a resolution of a possible violation of the second law of thermodynamics. The infall of a classical body into a black hole decreases the entropy of the system unless one can attribute an increase to the black hole entropy. Accordingly, the generalized second law of thermodynamics (GSL) states that the sum of the entropy of the black hole and matter fields can never decrease~\cite{Bekenstein1974}, i.e. $\delta\left(S_{BH}+S_m\right)\geq 0$.

Studying the infall of a small test body into the black hole, Bekenstein proposed~\cite{Bekenstein1981} an upper bound for the entropy-to-energy ratio of any physical system given by $ S\le (2\pi k_B/\hbar c ) \mathcal{ER}$. Subsequent works generalized this bound by including the angular momentum~\cite{Zaslavskii1991,Zaslavskii1992} and the charge of the system~\cite{Hod1999a,Hod1999b}. The most general bound reads~\cite{Bekenstein1999}
\be \label{eq:BekBound}
S\le\frac{2\pi k_B}{\hbar c}\left(\sqrt{\left(\mathcal{ER}\right)^2-c^2J^2}-\frac{q^2}{2}\right)\ ,
\en
where $\mathcal{R}$ is the radius of the minimum sphere that encloses the system, $\mathcal{E}$ is the energy, $J$ is the angular momentum, and $q$ is the charge. The above relation is assumed to be valid for an arbitrary physical system and, indeed, it has been confirmed in a variety of physical situations~\cite{Sorkin1981, Bekenstein1984, Bekenstein1989, Schiffer1990, Zaslavskii1996}. Furthermore, the equality in \eqref{eq:BekBound} is achieved only for the Kerr-Newman black hole or the associated black hole of the physical situation, namely Reissner-Nordstr\"om (RN-BH) and Schwarzschild (SBH) for the static charged and neutral black hole, respectively. Most importantly, the entropy upper bound~\eqref{eq:BekBound} is assumed to be a universal bound, meaning that should be valid for arbitrary physical systems and independent on the underlying dynamics.

We shall analyze the entropy bound in the context of nonlinear electrodynamics (NLED). In particular, we consider the Born-Infeld (BI) electrodynamics~\cite{Born1933,Born410,Born425} that, among NLED, has several interesting features: avoidance of classical singularity, it emerges as the low-energy regime of string theory~\cite{Fradkin1985} and has no birefringence~\cite{Boillat1970,Plebanski1970} (see also \cite{Penafiel2017, Falciano2019} and references therein). We shall repeat the same thought experiment proposed by Bekenstein and collaborators of slowly lowering a test body into the black hole but now generalizing for a charged body obeying BI electrodynamics in the curved spacetime of an Einstein-Born-Infeld black hole (EBI-BH).

The paper is organized as follows. In the next section we briefly review NLED and show (see theorem \ref{thmNLEDSSBH}) that the displacement vector of a static charge in a static spherically symmetric spacetime is curl-free and specified by the maxwellian electrostatic potential in the same background. In section~\ref{BIBH} we describe the main properties of the EBI-BH and in \ref{TPABIBH} we solve the NLED to find the electrostatic potential of a BI test particle in the EBI-BH background. In section~\ref{Lowering} we calculate the change in the black hole area and show that Bekenstein's entropy bound is violated. We conclude with some final remarks in section~\ref{conclusion}.

\section{Nonlinear Electrodynamics}\label{NLED}

We define electromagnetism as a vector gauge theory for the $U(1)$ symmetry group where the Faraday tensor is given by $F_{\mu\nu}=\partial_{\mu}A_{\nu}-\partial_{\nu}A_{\mu}$. The dual of the Faraday tensor reads $\widetilde{F}^{\mu\nu}={1\over2}\eta^{\mu\nu\alpha\beta}F_{\alpha\beta}$ where $\eta^{\mu\nu\alpha\beta}$ is the totally antisymmetric Levi-Civita tensor. The electric and magnetic fields are defined as the projection along the normalized observer's worldline $v^\mu$, i.e $E^\mu=F^{\mu}_{\ \alpha}v^{\alpha}$ and $B^{\mu}=\widetilde{F}^{\mu}_{\ \alpha}v^{\alpha}$ such that $\ E^\mu v_\mu=B^\mu v_\mu=0$. There are only two linearly independent Lorentz invariants constructed with the Faraday tensor, its dual and the metric, namely, $	F\equiv{1\over2}F^{\mu\nu}F_{\mu\nu}=E_{\alpha}E^{\alpha}-B_{\alpha}B^\alpha$ and $	G\equiv {1\over2}\widetilde{F}^{\mu\nu}F_{\mu\nu}=2B_{\alpha}E^{\alpha}$. A generic NLED constructed solely in terms of these two invariants can be specified by its Lagrangian density $\mathcal{L}\left(F,G\right)$. In this case, the dynamics reads
\begin{eqnarray}\label{eq:fNLED}
\p_{\mu}(\sqrt{-g}E^{\mu\nu})=-4\pi\sqrt{-g}j^{\nu} \quad ,
\end{eqnarray}
where $E^{\mu\nu}$ and its dual $\widetilde{E}^{\mu\nu}=\frac12\eta^{\mu\nu\alpha\beta}E_{\alpha\beta}$, are the excitation tensor defined as
\begin{subequations}\label{Def:EET}
	\begin{eqnarray}
E^{\mu\nu}&=2\left(\mathcal{L}_FF^{\mu\nu}+\mathcal{L}_G\widetilde{F}^{\mu\nu}\right)\quad ,\\
\widetilde{E}^{\mu\nu}&=2\left(\mathcal{L}_F \widetilde{F}^{\mu\nu}-\mathcal{L}_G F^{\mu\nu}\right) \quad .
	\end{eqnarray}
\end{subequations}

The notation $\mathcal{L}_{X}$ means derivative of $\mathcal{L}$ with respect to $X$. There are also only two Lorentz invariant quantities constructed with $E^{\mu\nu}$ and $\widetilde{E}^{\mu\nu}$, namely
\begin{subequations}\label{Def:PS}
	\begin{eqnarray}
	P&={1\over2}E^{\mu\nu}E_{\mu\nu}=4\left(\mathcal{L}_{F}^2-\mathcal{L}_{G}^2\right)F+8\mathcal{L}_F\mathcal{L}_{G}G \ ,\\
	S&={1\over2}\widetilde{E}^{\mu\nu}E_{\mu\nu}=4\left(\mathcal{L}_{F}^2-\mathcal{L}_{G}^2\right)G-8\mathcal{L}_F\mathcal{L}_{G}F\ .
	\end{eqnarray}
\end{subequations}

The decomposition of the excitation tensor follows closely that of the Faraday tensor, namely $D^\mu=-E^{\mu}_{\ \alpha}v^{\alpha}$ and $H^{\mu}=-\widetilde{E}^{\mu}_{\ \alpha}v^{\alpha}$, which are  respectively the four-dimensional electric displacement and magnetic $H$-field. These tensors allow us to reformulate the NLED using the P-framework~\cite{Plebanski1970,Salazar1987,Bronnikov2001} by defining the hamiltonian-density
\begin{equation} \label{eq:Hden}
\mathcal{H}={1\over2}E_{\mu\nu}F^{\mu\nu}-\mathcal{L}=2\left(\mathcal{L}_{F}F+\mathcal{L}_{G}G\right)-\mathcal{L}\ .
\end{equation}

In order to complete the Legendre transformation we need to invert \eqref{Def:EET} to write $F^{\mu\nu}$ as a function of the excitation tensor and its dual. If that is the case, we can write
\begin{equation}\label{InvFE}
F^{\mu\nu}=
2\frac{\p\mathcal{H}}{\p E_{\mu\nu}}
=2\left(\mathcal{H}_{P}E^{\mu\nu}+\mathcal{H}_{S}\widetilde{E}^{\mu\nu} \right)\quad ,
\end{equation}
where $\mathcal{H}_{X}$ means derivative with respect to $X$. Then, the Lagrangian can be written in terms of the Hamiltonian as $\mathcal{L}(P,S)=2\left(\mathcal{H}_{P}P+\mathcal{H}_{S}S\right)-\mathcal{H}$. Note that even in NLED, the dynamics is linear in term of the excitation tensor. Nevertheless, due to the nonlinearity of the constitutive relations, $E^\mu$ and $B^\mu$ do not share the same symmetries and physical properties of $D^\mu$ and $H^\mu$. In particular, an electrostatic configuration implies $\mathbf{E}=-\boldsymbol{\nabla}\phi(x)$ but in general $\boldsymbol{\nabla}\times \mathbf{D}\neq 0$.

Notwithstanding, in a recent paper~\cite{Falciano2019}, it has been shown that the displacement vector of an arbitrary NLED in Schwarzschild spacetime is curl-free. In addition, assuming that the constitutive relations are invertible, the electrostatic potential solution can be written in terms of Linet's solution. The latter is the electrostatic potential for a test particle in conformity to Maxwell electrodynamics in Schwarzschild. We now generalize this result for an arbitrary static and spherically symmetric spacetime.

\begin{thm}\label{thmNLEDSSBH}
	The electrostatic potential $\phi(x)$ produced by a charged particle satisfying a generic NLED theory $\mathcal{L}(F,G)$ in a static spherically symmetric spacetime is entirely specified by the electrostatic potential $\psi(x)$ satisfying Maxwell's electromagnetism in the same background. The displacement vector is curl-free and given by $\mathbf{D}=-\boldsymbol{\nabla}\psi(x)$.
\end{thm}

\textit{Proof.}--- The electrostatic case implies $\mathbf{B}=0$ and $\partial_t\mathbf{E}=0$, hence the electric displacement reads $D^\mu=-2 \mathcal{L}_F(E)\, E^\mu$, where the electric field is given by the gradient of the potential $\phi(x)$, i.e. $E^\mu=\left(0, -\boldsymbol{\nabla}\phi \right)$ in the reference frame where the particle is at rest. Therefore, the dynamics~\eqref{eq:fNLED} reads
\begin{align}\label{NLEDisotropic}
	\p_\mu\left(\sqrt{-g}D^{\mu}\right)=-4\pi \sqrt{-g}\rho \quad ,
\end{align}
where the density is defined as $\rho=J_\alpha v^\alpha$, and we have used \eqref{Def:EET} together with the fact that the four-velocity of a static particle $v^\mu=c\, \delta^\mu{}_0/\sqrt{g_{00}}$ satisfies $\nabla_\mu v^\mu=\p_\mu v^\mu=0$. Let $\psi(x)$ be an auxiliary scalar function defined as the integral along the path with tangent vector $\mathbf{\dd l}$ such that 
\begin{align} \label{def:auxfunc}
	\psi ( x)&= -2 \int  \ \mathcal{L}_F(\boldsymbol{\nabla}\phi) \boldsymbol{\nabla}\phi \cdot \mathbf{\dd l} \quad  .
\end{align}
Straightforward inspection shows that \eqref{NLEDisotropic} is formally identical with the electrostatic Maxwellian case with the electric field given by the gradient of the function $\psi(x)$. Thus the displacement vector reads $\mathbf{D}=2 \mathcal{L}_F(\boldsymbol{\nabla}\phi)\, \boldsymbol{\nabla}\phi=-\boldsymbol{\nabla}\psi$. Furthermore, assuming that the constitutive relations are invertible, the P-framework allow us to write the Faraday tensor as a function of the excitation tensor and its dual. In the electrostatic case, \eqref{InvFE} shows that $\boldsymbol{\nabla}\phi= -2 \mathcal{H}_P(\boldsymbol{\nabla}\psi)\, \boldsymbol{\nabla}\psi$.  \hfill$\square$

In the present work we are interested in the physics of a BI charged test particle in the EBI-BH spacetime. The BI theory can be defined by the action
\begin{equation} \label{eq:LBI}
S=\frac{1}{4\pi}\int \dd^4 x\sqrt{-g} \ \beta^2\left(1-\sqrt{U}\right)\quad ,
\end{equation}
where $\beta $ is the parameter that specifies the maximum value of the field strength and $U=1+F/\beta^2-G^2/4\beta^4$. The BI theory is an example of a NLED whose $P$-framework is well-defined and completely analogous to the $F$-framework. Indeed, defining $V=1-P/\beta^2-S^2/4\beta^4$ one can show that $S=G$, $P=F-(F^2+G^2)/(U\beta^2)$ and $V= \left(1+{G^2}/{4\beta^4}\right)^2/U$.
Thus, the constitutive relations are
\begin{subequations} \label{BIfftilde}
	\begin{eqnarray}
F^{\mu\nu}&= -\frac{1}{\sqrt{V}}\left(E^{\mu\nu}+\frac{S}{2\beta^2} \widetilde{E}^{\mu\nu} \right)\quad ,\\
E^{\mu\nu}&=-\frac{1}{\sqrt{U}}\left(F^{\mu\nu}-\frac{G}{2\beta^2}\widetilde{F}^{\mu\nu}\right)\quad .
\end{eqnarray}
\end{subequations}

In particular, in the electrostatic case, the displacement vector simplifies to $\mathbf{D}= -2\mathcal{L}_F(E) \mathbf{E}$ and the above constitutive relations read
\begin{align}\label{constEqsESt}
\mathbf{D}= {\mathbf{E} \over{\sqrt{1- |\mathbf{E}|^2 \beta^{-2} }}}  &\quad \mbox{and}&
\mathbf{E}= {\mathbf{D} \over{\sqrt{1+ |\mathbf{D}|^2 \beta^{-2} }}} \ . 
\end{align}

\section{Einstein-Born-Infeld black hole}\label{BIBH}
The solution of a EBI-BH of charge $q$ and mass $M$ can be written in standard coordinate system $(t,r,\theta,\phi)$ as~\cite{Garcia1984,Breton2002,Breton2003,Breton2005,Chemissany2008,Linares2015,Gunasekaran2012,Fernando2006}
\begin{equation}\label{BImetric}
\dd s^2=g_{00}(r) c^2\dd t^2 -g_{00}^{-1} (r) \dd r^2 -r^2 \dd \Omega^2
\end{equation}
with $\dd \Omega^2=\dd \theta^2+\sin^2\theta\dd \phi^2 $. The time-time component of the metric is given by
\begin{align}
g_{00}(r)=1-\frac{r_s}{r}&+\frac{2r_q^2}{3r_\beta^4} \left(r^2-\sqrt{r^4+ r_\beta^4 }\right)\nonumber\\&+\frac{4r_q^2}{3r}\int_{r}^{\infty} \frac{\dd r'}{\sqrt{r'^4+r_\beta^4}} \quad ,
\label{BIg00}
\end{align} 
where $r_s\equiv 2G M/c^2$, $r_q=\sqrt{G}q/c^2$ and $r_\beta\equiv \sqrt{q/\beta}$. These three length parameters characterize completely the EBI-BH solution. Changing the limits of integration on the last term of \eqref{BIg00} to the complementary interval $\left(0,r\right)$ we obtain instead the Demianski's\cite{Demianski1986} particle-like solution (also referred as EBIon). The non-zero components of the electromagnetic field $F_{\mu\nu}$ and the excitation tensor $E_{\mu\nu}$ of the black hole are
\begin{align}
F_{01}(r)=\frac{q}{\sqrt{r^4+r_\beta^4}}&\quad  , &E_{01}(r)=\frac{q}{r^2}\quad .
\end{align}

\begin{center}
	\begin{figure}[t]
		\includegraphics[width=0.37\textwidth,height=0.25\textwidth]{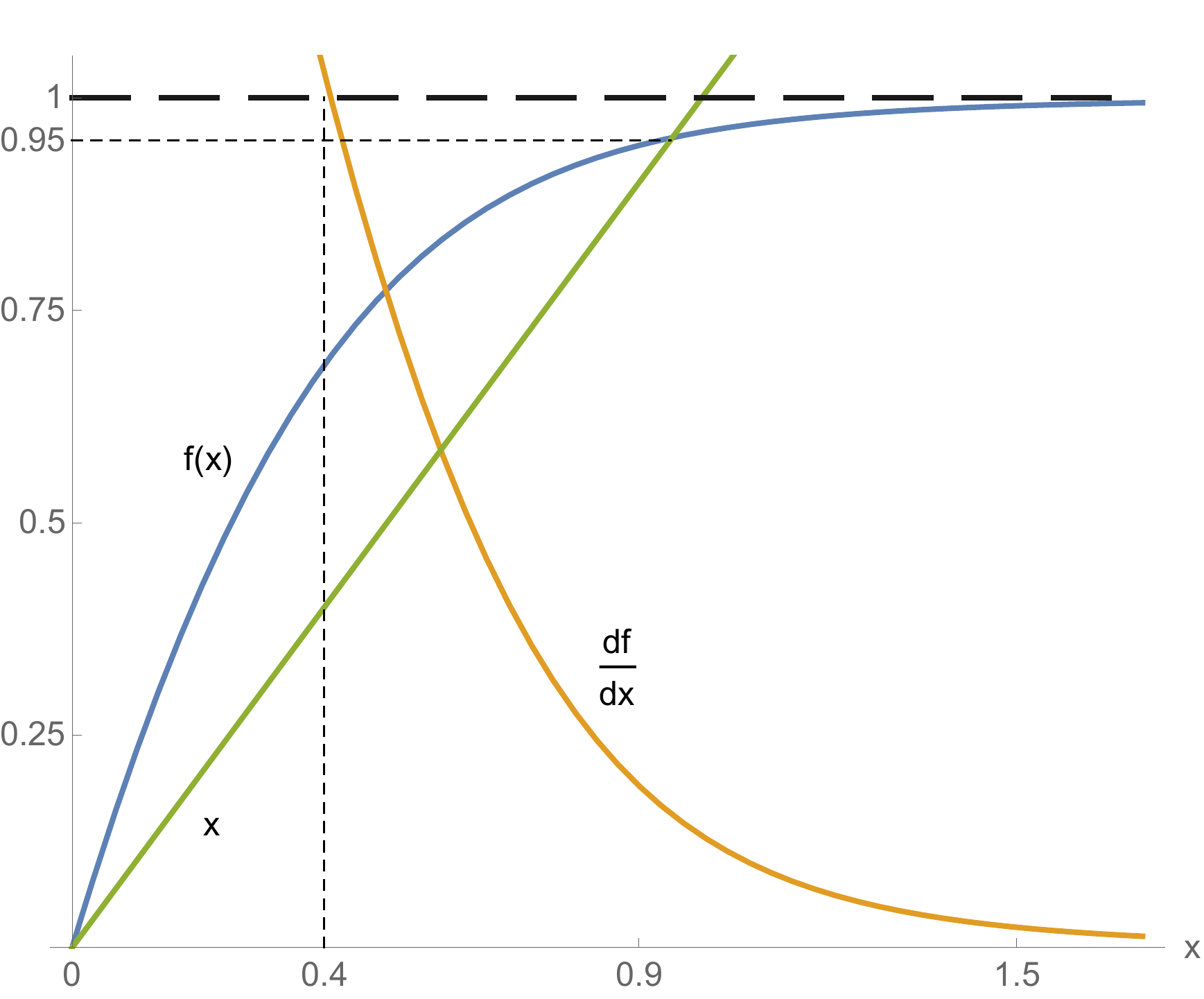}
		\caption{The function $f(x)$ defined in \eqref{BIg00f} is a monotonic increasing function that tends to 1 for large values of $x$. We can see that $f(x)>x$ for values smaller than $x\approx 0.95$. Its first derivative becomes smaller than $1$ for $x\approx 0.4$, while the second derivative is always negative for $x\in\left[0,\infty\right)$.}
		\label{fig:Function_f_x}
	\end{figure}
\end{center}

It is evident from the above expression that the displacement vector can be written as $\mathbf{D}=-\nabla\psi $ where $\psi(r)=q/r$ is the potential of an electric monopole at the origin. Notwithstanding, the potential for the electric field, $\mathbf{E}=-\nabla\phi_q$, reads 
\begin{align}\label{PhiBH}
\phi_q(r)&=q\int_{r}^{\infty} \frac{\dd r'}{\sqrt{r'^4+r_\beta^4}}=\frac{q}{r}\ {}_2F_1 \left[\frac14,\frac12,\frac54,-\frac{r_\beta^4}{r^4}\right]\quad ,
\end{align} 
where ${}_2F_1[a,b;c;z]$ is the Gaussian hypergeometric function. This is an instantiation of the theorem~\ref{thmNLEDSSBH} valid for any NLED in static spherically symmetric spacetimes. We can rewrite the metric in a more suggestive way as
\begin{align}
g_{00}\left(r\right)=&1-\frac{r_s}{r}+\frac{r_q^2}{r^2}f\left(\frac{r}{r_\beta}\right) \label{BIg00f}\\
f\left(x\right)=&
\frac{2}{3} x^4\left(1-\sqrt{1+\frac{1}{x^4}}\right)
+\frac{4}{3}\ {}_2F_1 \left[\frac14,\frac12,\frac54,-\frac{1}{x^4}\right] \nonumber 
\end{align} 

The function $f\left(x\right)$ is a monotonic increasing function that start at zero and approaches 1 as its argument goes to infinity (see figure~\ref{fig:Function_f_x}). Let us analyze the properties of the EBI-BH metric. For that, it is convenient to rescale the radial coordinate in order to absorb $r_\beta$ and define an adimensional quantity, namely $x=r/r_\beta$. Thus,
\begin{align}
g_{00}\left(x\right)=&1-\frac{1}{x}\Big(x_s-x_q^2 F\left(x\right)\Big) \label{BIg00fx}
\end{align} 
where $F(x)= f(x)/x$ and we redefined the parameters as $x_s\equiv  r_s / r_\beta$ and $x_q\equiv r_q / r_\beta$. The function $F(x)$ is also monotonic but now a decreasing function with $F(0)={\Gamma \left(\frac{1}{4}\right)^2}/{ 3\sqrt{\pi }}$ and $F(x\rightarrow\infty)=0$, hence if $x_s> x_q^2\, F(0)$ the metric diverges to minus infinity at the origin, i.e. $g_{00}\left(0\right)=-\infty$. Contrarily, if $x_s<x_q^2\, F(0)$, we have $g_{00}\left(0\right)=\infty$. Note that at spatial infinity $g_{00}\left(x \rightarrow\infty\right)=1$ irrespective of the values of the parameters (the EBI-BH is asymptotically flat). The first derivative of the metric component reads
\begin{align}
\frac{\dd g_{00}}{\dd x}\left(x\right)
=&\frac{1}{x^2}\left(x_s- x_q^2 \Big[F(x)-x\frac{\dd F}{\dd x}(x)\Big] \right) \label{BIdg00fx}\\
=&\frac{1}{x}\left(1-g_{00}(x)+2 x_q^2 \left[ x^2-\sqrt{1+x^4}\right] \right) \ .   \nonumber 
\end{align}

\begin{center}
	\begin{figure}
		\includegraphics[width=0.4\textwidth]{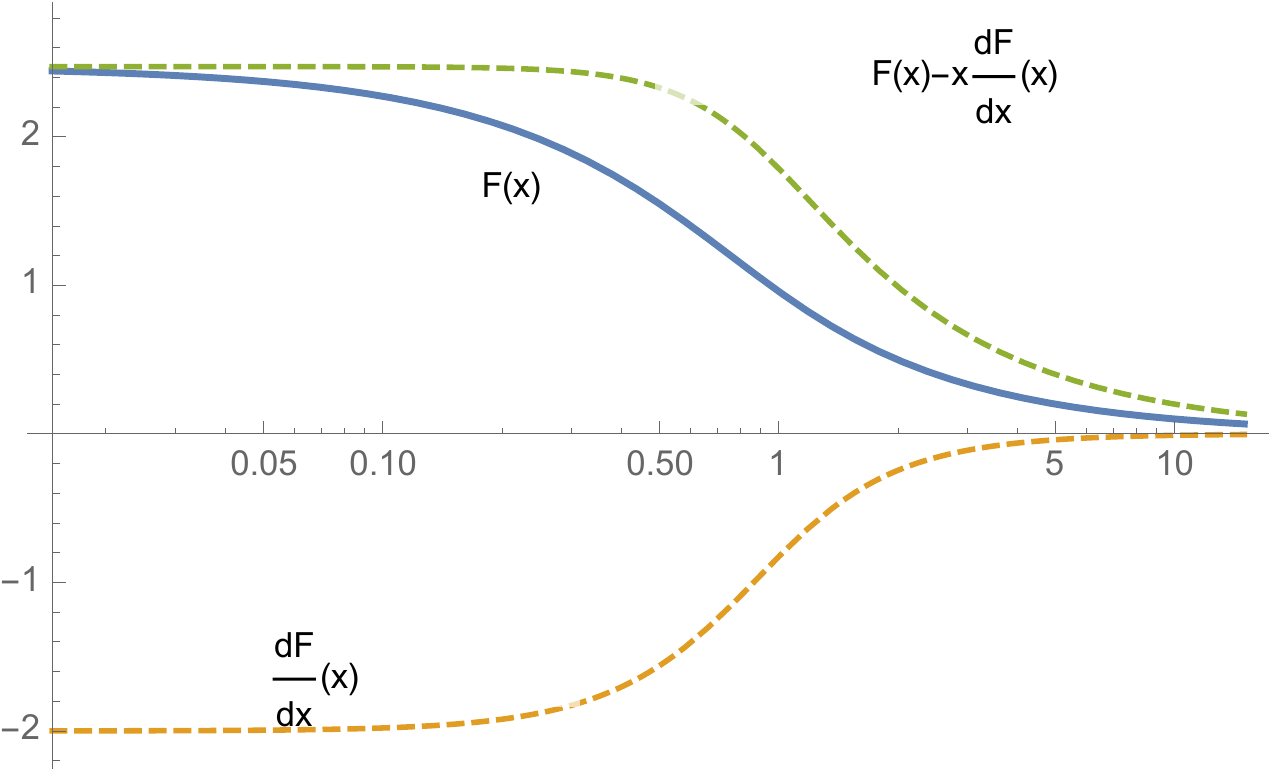}
		\caption{Plot of the function $F(x)$ its derivative and the particular combination of them that appears in the derivative of $g_{00}(x)$. All of them are monotonic function }
		\label{fig:Function_F_dF}
	\end{figure}
\end{center}

The function $\frac{\dd F}{\dd x}(x)$ is a monotonically increasing function starting at $-2$ and it approaches zero for $x\rightarrow\infty$. On the other hand, the sum $F\left(x\right)-x\frac{\dd F}{\dd x}(x)$ appearing in the derivative of the metric is monotonically decreasing and with the same limits as $F(x)$. It is $F(0)$ for $x=0$ and tends to zero for $x\rightarrow\infty $ (see figure~\ref{fig:Function_F_dF}). 

We conclude that if $x_s >x_q^2\, F(0)$, then $g_{00}(x)$ has no extreme point and there is only one horizon, since the $g_{00}(x)$ will inevitably cross zero (Schwarzschild-like black hole). Alternatively, if $x_s <x_q^2\, F(0)$ the $g_{00}(x)$ diverges to plus infinity at the origin and $g_{00}(x)$ has one extreme point. In this situation the black hole can have zero, one or two horizons (Reissner-Nordstr\"om-like black hole). Indeed the second derivative of the metric component reads
\begin{align}
	\frac{\dd^2 g_{00}}{\dd x^2}\left(x\right) &=-\frac{2}{x}\frac{\dd  g_{00}}{\dd x}\left(x\right)
	+4x_q^2 \left(1 - \frac{x^2}{\sqrt{1 + x^4}}\right)\ .  \label{BIddg00fx}
\end{align}
Thus, if $\frac{\dd  g_{00}}{\dd x}\left(x\right)=0$, the second derivative is necessarily positive that characterizes a minimum. There is, however, an extra condition for the appearance of two horizons; we need $g_{00}(x)<0$ at the minimum. Suppose there is a single horizon, namely, a position $x_{ext}$ such that $\frac{\dd  g_{00}}{\dd x}\left(x_{ext}\right)=g_{00}(x_{ext})=0$, i.e.
\begin{align}
1+x_q^2\frac{\dd F}{\dd x}(x_{ext})=0&\ \Rightarrow \ x_{ext}^2= x_q^2 -\frac{1}{4 x_q^2} \quad .
\end{align}

Since $x_{ext}^2 > 0$, the existence of an extremum black hole requires that $ x_q^2>1/2$. By definition, the parameters of the extreme EBI-BH are related by $x_s^2=4x_q^2 f\left(x_{ext}\right)$. Combining all the above relations, we find that the condition for a EBI-BH with two horizons is
\begin{align*} 
 \frac{\sqrt{36\pi }}{\Gamma \left(\frac14\right)^2}\frac{M}{\sqrt{q \beta}} < q  < \sqrt{\frac{G}{f\left(x_{ext}\right)}}M  \quad  , \quad 
\frac{c^4}{2G\beta} < q \quad  .
\end{align*}

In the limit $\beta\rightarrow \infty$, we re-obtain the RN condition $0<q<\sqrt{G}M$. For finite values of $\beta$ we can recast the above inequalities as
\begin{align}
  \frac{ \sqrt{18\pi} }{\Gamma\left(1/4\right)^{2}}  & < \frac{x_q}{x_s} < \frac{1}{\sqrt{4f\left(x_{ext}\right)}} \quad , \ 
&\frac{\sqrt{2}}{2} \leq  x_q \quad . \label{rangeqbeta}
\end{align}

Note that the above inequality for $r_q$ can be satisfied only if $f\left(x_{ext}\right)< \Gamma \left(\frac14\right)^4/ (72\pi)\approx 0.764$. The existence of two horizons imposes an upper bound on $f\left(x_{ext}\right)$.

We can also analyze the behavior of the metric in the two asymptotic regimes of the parameter $\beta$. As mentioned before, the metric \eqref{BImetric} is always asymptotically flat but it differs for small values of $r$ depending on the magnitude of $\beta$. In the limit $\beta\ll 1$ we have the deep BI regime where $r_\beta\gg1$ and we can approximate the time-time component of the metric by
\begin{align}\label{deepBImetric}
& g_{00}\left(r\right)=\left(1-2\frac{r_q^2}{r_\beta^2} 
\right)-\frac{r_{qs}}{r} +\mathcal{O}\left(\beta^{2}\right)\quad ,\nonumber \\
& \mbox{with} \quad r_{qs}=r_s\left(1-\frac{\Gamma \left(\frac{1}{4}\right)^2}{3 \sqrt{\pi }}\frac{r_q^2}{ r_s r_\beta}\right)\quad .
\end{align}

At first order in $\beta$, the BI corrections only modify the value of the Schwarzschild mass. For low values of $\beta$, the nonlinearities screen the charges and make them inefficient. As a consequence, the metric departs from RN-BH and behaves as SBH but with a rescaled mass. The horizon is located at $r_h\approx r_{qs}$.

The opposite regime happens in the limit $\beta\gg 1$. This correspond to the superficial BI regime where $r_\beta\ll 1$, and we can approximate the time-time component of the metric by
\begin{align}
g_{00}\left(x\right)=&1-\frac{x_s}{x} +\frac{x_q^2}{x^2}-\frac{x_q^2 }{20 x^6}+O\left(\frac{x_q^2}{x^{10}}\right)\quad .
\end{align}

This expansion corresponds simultaneously to the limit $r\rightarrow \infty $ with finite $\beta$ or $\beta\rightarrow \infty $ with finite $r$. The latter corresponds to the Maxwellian limit, hence it comes with no surprise that the metric reduces to RN at first order. Since $r_\beta\ll 1$, outside the external horizon, i.e. $r>r_q$, the term $x_q^2/x^6$ already gives only small corrections to the metric component. Therefore, in this case, the horizon should be located close to the external RN horizon $x_h\approx x_+$ with $x_+=1/2\left(x_s+\sqrt{x_s^2-4x_q^2}\right)$.

The horizon radius $x_h$ has no analytic solution and it can only be determined numerically through the equation $g_{00}(x_h)=0$. Nevertheless, we have an implicit solution given by
\begin{equation}\label{BIhorizonr}
x_{h}=\frac{1}{2}\left(x_s+\sqrt{ x_s^2 -4x_q^2f\left(x_{h}\right)} \right)\quad .
\end{equation}

The fact that $f(x)$ is a monotonic increasing function of $x$ provides us with an iterative procedure to obtain the location of the horizon. Let us define the first approximate solution by $x_{1}=1/2\left(x_s +\sqrt{x_s^2 -4x_q^2f (x_+)} \right)$. Note that, since $x_s>x_+>x_s/2$ and $1\geq f(x)\geq 0$, we have the following ordering $x_+<x_{h}<x_1<x_s$. 

To improve the solution we can substitute $x_+$ by $x_1$ in the argument of $f(x)$ defining the second radius as $x_{2}=1/2\left(x_s +\sqrt{x_s^2 -4x_q^2f (x_{1})} \right)$. Now $x_2<x_{h}$ and the ordering becomes $x_+<x_2<x_{h}<x_1<x_s$. This iterative procedure defines the radius $x_{k+1}$ as
\begin{equation}\label{BIhorizonri}
x_{k+1}=\frac{1}{2}\left( x_s+\sqrt{x_s^2-4x_q^2f\left(x_{k} \right)} \right)\ \mbox{with}\quad k\in \mathbb{N}\ .
\end{equation}

For every $k$ even (odd), we have $x_k$ smaller (greater) than $x_h$. Therefore the sequence of radius can be ordered as
\[
x_+<x_2<x_4 <\dots<x_{h}<\dots  <x_3<x_1<x_s \ .
\]

In the limit $k\rightarrow\infty$ the sequence approaches the true value $x_h$ but even for the first radius we already have a good approximate solution. In order to check this, let us evaluate the error in truncating the above sequence. By definition, the horizon radius satisfies \eqref{BIhorizonr}. Moreover, the condition $r_\beta\ll 1$ together with the relation $x_s/2\leq x_+ \leq x_{h}$ imply that $x_h\approx x_+\gg 1$ and $1\geq f\left(x_{h} \right)\geq f\left(x_+ \right) \geq f\left(1\right)\approx 0.96$. Therefore, the error in placing $x_+$ instead of $x_h$ in the argument of the function $f(x)$ is smaller than $\delta f\sim 0.04$. Taylor expanding we find
\begin{equation}\label{Taylorf}
r_{h}\approx r_1-0.04 \frac{r_q^2}{r_s}+\mathcal{O}\left(\delta f^2\right)\quad .
\end{equation}

\begin{figure}[t]
	\includegraphics[width=0.47\textwidth,height=0.23\textwidth]{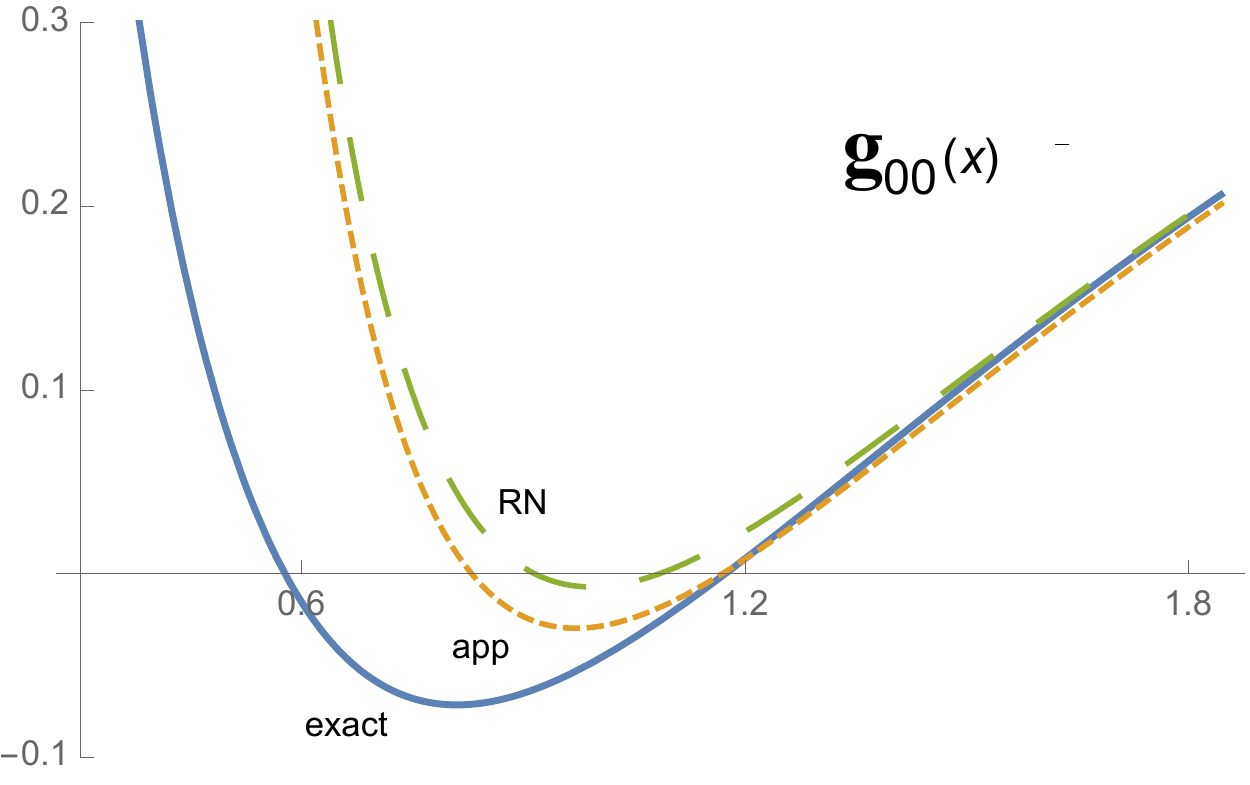}
	\caption{Plot of the time-time component of the metric tensor as a function of the adimensional variable $x=r/r_\beta$ with $x_s=2$ and $x_{q}=0.996$. The solid curve is the exact function $g_{00}(x)$  (see \eqref{BIg00fx}). The small dashed curve displays the approximate solution given by \eqref{g00appox} where the function $f(x)$ is fixed at its horizon value. The large dashed shows $g_{00}(x)$ for the RN-BH. We see that our approximate solution is very different inside the exterior horizon but can be considered as a good approximation at and beyond the horizon. As expected the location of the RN horizon is smaller than the BI horizon and the three $g_{00}$ are similar far away from the black hole.}
	\label{fig:metric_approx}
\end{figure}

The function $f(x)$ becomes almost constant for large values of $x$, hence it should not vary appreciably outside the exterior horizon. Therefore, we can further simplify the metric by freezing the value of this function at the horizon, i.e.
\begin{align}\label{g00appox}
g_{00}\left(x\right) \approx &1-\frac{x_s}{x} +\frac{x_q^2f(x_{h}) }{x^2}\quad \mbox{for}\ x>x_h\ .
\end{align}

Figure~\ref{fig:metric_approx} compares the exact form~\eqref{BIg00fx} of $g_{00}(x)$ with the approximative solution~\eqref{g00appox} and the RN-BH metric. Even though inside the horizon the approximation fails completely, at the external horizon and beyond, it is a good approximation to fix the value of the function at $f(x_h)$. Moreover, the rescale of the black hole charge as $q^2f(x_h)$ is crucial to describe the metric close to the horizon. This plot shows that the RN-BH metric does not reproduce the EBI-BH close the exterior horizon (only in the limit $\beta\rightarrow\infty$).

As a last comment, we mention that there is an upper and lower bound for the size of the horizon area. The time-time component of the metric satisfies
\begin{align}
	1-\frac{r_s}{r}\leq  g_{00}\left(r\right) \leq 1-\frac{r_s}{r}+\frac{r_q^2}{r^2} \quad ,
\end{align}
and this relation can be translated into area by stating that the EBI-BH area $(A_{BI})$ is always smaller than SBH and greater than the exterior RN-BH area, i.e.
\begin{equation}
4\pi r_+^2\leq A\left(r_{BI}\right) \leq  4\pi r_s^2 \quad ,
\end{equation}
where $2r_{\pm}=r_s \pm\sqrt{r_s^2-4r_q^2}$ and equality holds for the two limiting behaviors $\beta\rightarrow\infty$ or $0$.

\section{Charged particle in the vicinity of a EBI-BH}\label{TPABIBH}

Let us consider a test charged body with mass $m$ and charge $e$. In order to find the electrostatic potential associated with this test body, we need to solve the NLED for a static test particle in the black hole spacetime. This becomes a straightforward exercise by using theorem~\ref{thmNLEDSSBH}. The challenge rests, then, in solving Maxwell's equation in the same spacetime. 

Copson and then Linet~\cite{Copson1928, Linet1976} obtained the solution for the electrostatic potential of a test particle in SBH, which later was generalized for RN-BH~\cite{Leaute1976}. Our analysis of the preceding section shows that for our purpose we can describe the EBI-BH as a RN-BH with a rescaled charge given by $q^2\rightarrow q^2f(x_h)$ (see~\eqref{g00appox}). Therefore, we can use Linet's solution with a few adjustments to describe the electrostatic potential of a test particle outside the exterior horizon of a EBI-BH. In this section we summarize the main steps to obtain the electrostatic solution for RN from the solution for SBH.

Consider a static spherically symmetric spacetime. By choosing the time coordinate along the integral lines of the timelike Killing vector $\mathbf{\xi}_t=\p_t$ and the third coordinate along the angular Killing vector $\mathbf{\xi}_\phi=\p_\phi$, namely a coordinate system $(t,r,\theta, \phi)$, the metric does not depends on $(t,\phi)$. The dynamics of a test charged particle reads
\begin{align}\label{eq.Eab}
\p_{\mu}(\sqrt{-g}g^{\mu\alpha}g^{\nu\beta}E_{\alpha\beta})=-4\pi\sqrt{-g}j^{\nu} \quad ,
\end{align}
where $\mathcal{L}_{\xi}  g_{\mu\nu}=0$. In the static case, there is no magnetic field, hence we have $\mathbf{D}=-2\mathcal{L}(E)\mathbf{E}$. By symmetry arguments, the excitation tensor also satisfies $\mathcal{L}_{\xi}  E_{\mu\nu}=0$, hence we have $\mathbf{D}=\mathbf{D}\left(r,\theta\right)$. The displacement vector of a NLED satisfies the same differential equation of the electric field for the Maxwellian case, which is given as the gradient of the electric potential. Thus, we can write $\mathbf{D}=-\nabla \psi (r,\theta)$ and  \eqref{eq.Eab} becomes
\begin{align}\label{eq.psi}
	\frac{1}{r^2}\frac{\p }{\p r}\left(r^2\frac{\p \psi}{\p r}\right)+\frac{g^{00}(r)}{r^2}\hat{L}^2\psi=-4\pi j^0	\quad ,
\end{align}
where $\hat{L}^2$ is the square angular-momentum operator
\[
\hat{L}^2\psi=
\frac{1}{\sin\theta}\frac{\p }{\p \theta}\left( \sin \theta \frac{\p \psi}{\p \theta}\right)
+\frac{1}{\sin^2 \theta}\frac{\p^2 \psi}{\p \varphi^2} \quad .
\]

The source is considered as a point-like particle located at $r=a$ and $\cos \theta =1$, hence $j^0= e a^{-2}\delta \left(r-a\right)\delta \left(\theta\right)$. The case of a charged particle with charge $e$ outside a SBH, where $g_{00}(r)=1-r_s/r$, has been studied by Copson and Whitaker~\cite{Copson1928}. However, their solution does not have the correct asymptotic behavior. Linet showed that Copson's solution in fact describes two charges. Analyzing the limit $r\rightarrow \infty$ it appears a charge $e$ at $\left(r=a,\theta =0\right)$ and another charge $-er_s/2a$ inside the black hole. In order to correct the solution Linet included a spherically symmetric term of the form $er_s / 2a r$.

The RN case can be written in terms of the above solution by a coordinate transformation that maps the differential equation for the electrostatic potential in RN-BH into the differential equation for SBH.  This procedure works only if the metric can be written as $g_{00}=(1-r_+/r)(1-r_-/r)$ and the difference $r_+-r_-$ is independent of $r$. 

There is little hope to find a similar coordinate transformation to map the electrostatic equation in EBI-BH into SBH. Indeed, the EBI-BH metric cannot be written as a polynomial and has a nontrivial dependence on $r$ through the function $f(x)$ defined in \eqref{BIg00f}. Notwithstanding, outside the exterior horizon, the BI corrections only suppress the effect of the black hole charge and the metric can be approximated by fixing the function $f(x)$ at the exterior horizon as proposed in \eqref{g00appox}. With this approximation we have 
\begin{align}
g_{00}(r)
&=\frac{\left(r-r_h\right)\left(r-r_\ast\right)}{r^2}\quad ,
\end{align}
where $r_h=1/2 \left(r_s+\sqrt{r_s^2-4r_q^2f(x_h)}\right)$ and $r_\ast=r_s-r_h$. Straightforward substitution shows that the coordinate transformation $z=r-r_\ast $ together with a redefinition of the potential as 
\begin{align}
 \psi\left(r,\theta\right)&=\frac{z}{z+r_\ast}\psi_s\left(z,\theta\right)
\end{align}
brings the equation EBI-BH~\eqref{eq.psi} to the form
\begin{align}\label{eq.psi2}
\frac{\p }{\p z}\left(z^2\frac{\p \psi_{s}}{\p z}\right)+ \left(1- \frac{r_h-r_\ast }{z}\right)^{-1} \hat{L}^2 \psi_{s}=-4\pi z\left(z+r_\ast\right) j^0	\quad ,
\end{align}
which is formally identical
\footnote{The right hand-side is not strictly identical since it is not $-4\pi z^2$. Notwithstanding, the source is a delta function which trivializes the necessary modification in the solution.}
to the equation for the electrostatic potential in SBH with the substitution $r_s^2\rightarrow r_s^2-4r_q^2f(x_h)$. Using the derivation of appendix~\ref{secappARN}, the electrostatic potential in the vicinity of a EBI-BH reads
\begin{widetext}
	\begin{align}\label{LinetBI}
\psi(r,\theta)=\frac{e}{2ar}\frac{(2a- r_s )(2r- r_s)- \left(r_s^2-4r_q^2f(x_h)\right)\cos\theta }{\left[\left(2r- r_s \right)^2+\left(2a-r_s \right)^2-2(2a- r_s)\left(2r- r_s \right)\cos \theta-\left(r_s^2-4r_q^2f(x_h)\right)\sin^2\theta\right]^{1/2}}+\frac{er_s}{2ar}\quad ,
	\end{align}
\end{widetext}
where the last term is Linet's correction of the asymptotic behavior (see discussion above \eqref{app:LinetRN}). The above expression gives the potential in standard coordinates but, for future analysis, it is convenient to transform it to isotropic coordinates. In the isotropic coordinate system, the interval reads
\begin{align}
\dd s^2=&\frac{ \left(1-\aiso \biso_q / \riso^2 \right)^2}{\Sigma^2} c^2\dd t^2-\Sigma^2\left[\dd \riso^2+\riso^2\dd \Omega^2\right]\ ,
\end{align}
where $\aiso$ is the position of the charged particle in the isotropic coordinate system and
\begin{align}
\Sigma(\riso)=\left(1+\frac{r_s}{4\riso}\right)^2-\frac{r_q^2f(x_h)}{4\riso^2} & , &\biso_q=\frac{r_s^2-4r_q^2f(x_h)}{16\aiso}\quad . \label{def:SigmaRN}
\end{align}

Using this coordinates, Linet's solution reads (see appendix~\ref{secappIsoCoord} for details)
\begin{align}
\psi(\riso,\theta)&=\psi_{Cop}(\riso,\theta)+\psi_p(\riso)\label{psiLinet}
\end{align}
where $\psi_p(\riso)$ is the term added by Linet, which in isotropic coordinates reads $\psi_p(\riso)= er_s  (\aiso \Sigma_{\aiso})^{-1}/ 2(\riso  \Sigma)$, with $\Sigma_{\aiso}\equiv \Sigma(\aiso)$, and 
\begin{align}
\psi_{Cop}(\riso,\theta)
&=\frac{e}{\riso}\frac{\Sigma^{-1}_{\aiso}}{\Sigma}
\frac{\left( \mu^2(\riso,\theta)+\biso_q/\aiso\right)}{\mu(\riso,\theta)}
\label{psiCop}\\
\mu(\riso,\theta)&=\sqrt{\frac{\riso^2+\biso_q^2-2\biso_q\riso\cos\theta}{\riso^2+\aiso^2-2\aiso\riso\cos\theta}} \quad . \label{muabBI}
\end{align}

Given the above solution, we can now apply theorem \ref{thmNLEDSSBH}. The constitutive relations for BI electrodynamics gives
\be\label{constitBI2}
\mathbf{E}(x)=-\boldsymbol{\nabla}\phi(x)=-\frac{\boldsymbol{\nabla}\psi(x)}{\sqrt{1+|\boldsymbol{\nabla}\psi(x)|^2\beta^{-2}}}\ .
\en
In principle one can integrate \eqref{constitBI2} and find the exact electrostatic potential $\phi(x)$ as a function of $\psi(x)$ and $\boldsymbol{\nabla}\psi(x)$. But in practice, one obtains only a numerical $\phi(x)$ since the above equation has no analytical solution. In~\cite{Falciano2019}, we have shown that a good approximation is to split the electric field as a sum of two terms
\be\label{aproxBIVcVp}
\mathbf{E}(x) \approx-\frac{\boldsymbol{\nabla} \psi_{Cop}}{\sqrt{1+|\boldsymbol{\nabla} \psi_{Cop}|^2\beta^{-2}}}-\boldsymbol\nabla \psi_{p} \ ,
\en
which is equivalent to decompose the BI electrostatic potential as
\be \label{eq:phiapproximation}
\phi(x)\approx \phi_{Cop}(x) +\psi_{p}(x) \ .
\en

The first term $\phi_{Cop}(x)$ is obtained by integrating \eqref{constitBI2} using only $\psi_{Cop}(x)$ in place of $\psi(x)$, namely, it is the BI solution if one considers only Copson's  solution for the Maxwellian dynamics. The $\psi_{p}(x)$ represents the added term to correct the asymptotic behavior of Copson's solution. As discussed in \cite{Falciano2019}, $\psi_{p}(x)$ encodes the black hole horizon polarization due to the presence of the test charged particle. Alternatively, one can interpret this extra term as an image charge inside the black hole. 

Since the horizon works as a conducting surface~\cite{Price1986, MacDonald1985}, the BI nonlinearities of the image particle must remain confined inside the horizon. Therefore, viewed from the outside, it is a very good approximation to consider the image particle as a Maxwellian charge. This also implies that any NLED with the correct Maxwellian limit for weak fields must produce the same black hole polarization.

\begin{figure}[t]
\includegraphics[width=0.47\textwidth,height=0.28\textwidth]{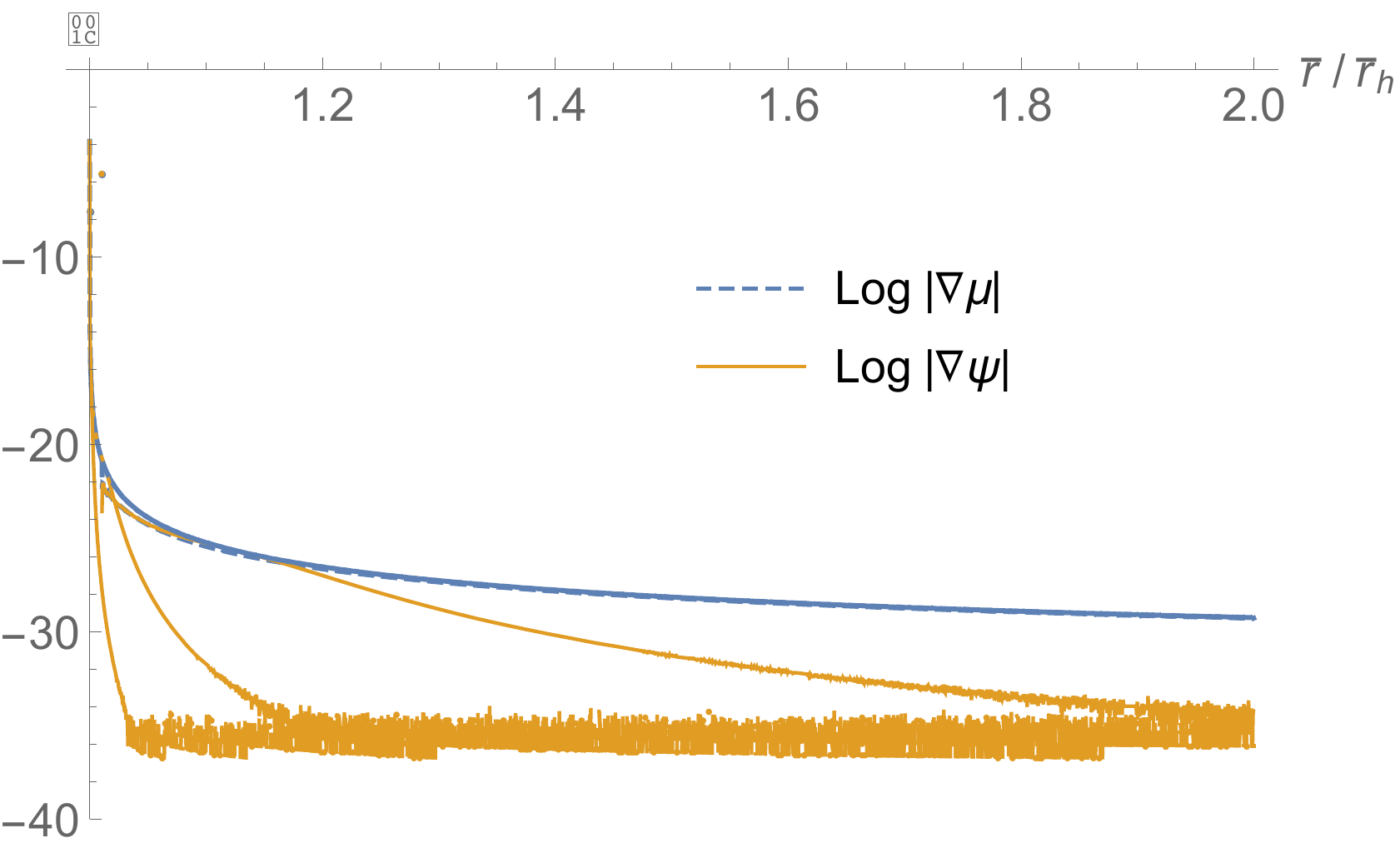}
\caption{We plot the relative error in approximating the gradient of $\mu$ and $\Psi$ using \eqref{gradPsiapprox}. We plot in log scale the difference of the exact and the approximated expression divide by the exact value. The exact values are calculated by numerically integrating $|\boldsymbol{\nabla} \mu |$ and $|\boldsymbol{\nabla}_y\Psi |$ from infinity along the $\theta=0$ axis where the particle is located at $\aiso$. We plot the relative error for three locations of the charge particle, i.e. $\aiso/\riso_h-1=\left( 10^{-2},\, 10^{-4}, \, 10^{-6}\right)$.
}
\label{fig:test_approxs}
\end{figure}

We can extract the dependence of the electrostatic potential on the charges and the field's strength parameter $\beta$ by defining dimensionless coordinates. Similar to $r_\beta$, we can define a characteristic length associated to the charge $e$ as $r_e\equiv \sqrt{e/\beta}$. In this manner, we define the dimensionless isotropic radius $\yiso\equiv \riso/r_e$, and its related quantities $\yiso_a=\aiso/r_e$, $\yiso_{b}=\biso_{q}/r_e$, $y_s=r_s/r_e$, $y_q=r_q/r_e$, and so forth. The gradient scales as $\boldsymbol{\nabla}=r_e^{-1}\boldsymbol{\nabla}_y$ and the dimensionless Linet's potential is defined as $\psi(x)=\sqrt{e\beta}\, \Psi(y)$. In dimensionless quantities, \eqref{constitBI2} becomes
\be\label{dimlessBI}
\boldsymbol{\nabla}_y\Phi(y)=\frac{\boldsymbol{\nabla}_y\Psi(y)}{\sqrt{1+|\boldsymbol{\nabla}_y\Psi(y)|^2}}\ , 
\en
where $\Phi(y)\equiv \frac{1}{\sqrt{e\beta}}\phi(x)$. In order to integrate \eqref{dimlessBI} we need to tame the divergences of $\mu(y,\theta)$ and $\Psi(y)$ as one approaches the charged particle. Far away we have $\mu\rightarrow 1$ and $\Psi\rightarrow 1/y$, but close to the charged particle $\mu\rightarrow \infty$ and $\Psi\rightarrow \Psi_c\approx \mu\Sigma^{-2}/y_a$. We can incorporate both behaviors by approximating (see figure~\ref{fig:test_approxs} and appendix~\ref{app:mathapprox}) 
\begin{align}
|\boldsymbol{\nabla} \mu |  &\approx \mu^2\frac{r_e(\aiso-\biso_q)}{(\riso-\biso_q)^2\Sigma(\riso)} 
\quad , \quad |\boldsymbol{\nabla}_y\Psi | \approx  \left(\frac{\Psi_{Cop}}{\zeta}\right)^2 \label{gradPsiapprox}
\end{align}
with 
\begin{align}
\zeta^{-2}(\riso,\theta)&=\Sigma(\aiso) \frac{ \left( \mu^2-\biso/\aiso\right)}{\left(\mu+\biso/\aiso\mu\right)^2}\Bigg[\frac{\riso (\aiso-\biso)}{(\riso -\biso)^2} \label{defzeta}\\
&\quad +\frac{\left(\riso^2-\aiso \biso \right) }{\riso^2 \Sigma(\riso)}\frac{\left( \mu^2+\biso/\aiso\right) + \mu r_s/2\aiso }{\mu  \left( \mu^2-\biso/\aiso \right)}\Bigg] \nonumber
\end{align}

The function $\zeta(\riso,\theta)$ has two limits given by $\zeta_{\aiso}\equiv \zeta(\aiso,0)=\sqrt{ \left(1-\biso/\aiso\right)/\Sigma(\aiso)}$ and $\zeta_\infty\equiv \zeta(\infty,\theta)=
(1+\biso/\aiso)/\Sigma (\aiso) $. The variability of $\zeta$ prevent us from integrating \eqref{dimlessBI}. We need a further simplification. The simplest solution is to fix $\zeta$, which must be taken at the particle's location in order to maintain the normalization $|\boldsymbol{\nabla}_y\Phi (\aiso)|=1$. This is already a very good approximation if the particle is not too close to the horizon. At the black hole horizon, the BI-region\footnote{When $|\boldsymbol{\nabla}_y\Psi|= \lambda^2 $, with $\lambda\geq 2$, the BI field strength is approximately $\lambda^2$ times lower than the Maxwellian strength. The exact value is not important as long as we keep it of order unit.}, defined by $|\boldsymbol{\nabla}_y\Psi |>\lambda^2$ with $\lambda \gtrsim 2$, deforms into an oval spheroid (see figure 4 of \cite{Falciano2019}). As a consequence the transition from a Maxwell-like behavior to a BI potential becomes steeper.

In any case, when the BI nonlinearities dominate, the electric field is almost constant to $|\boldsymbol{\nabla}_y\Phi |\approx1$. In this case, we can decompose the system in two regions that are matched at the border of the BI region, which we specify by $\yiso_{\ast}$. Outside the BI-region, the potential is almost maxwellian, $\Phi \approx \Psi$, while inside the BI-region the potential becomes an increasing linear function $\Phi(\yiso)=\Phi(\yiso_{\ast}) +\Sigma_{a}(\yiso_{\ast}-\yiso)$. We have used normalized variable to show that the above approximations are valid irrespective of the value of the BI field strength parameter $\beta$. Now we can come back to the isotropic coordinates \eqref{def:SigmaRN}. The position of the transition into the BI-region $\yiso_{\ast}$ depends on the parameter $\beta$ but $\Phi(\yiso_{\ast})$ does not since it is by definition the potential from the Maxwellian region. The transformation from $\phi$ to $\Phi$ gets a factor $\sqrt{e\beta}$ while $\yiso_{\ast}=\riso_{\ast}/r_e$, hence we see that $\phi_{Cop}(\aiso) =\sqrt{e\beta}\psi(\riso_{\ast}) +\beta\left(\riso_{\ast}-\aiso\right)$. The radius $\riso_{\ast}$ can be calculated by solving $|\boldsymbol{\nabla}_y\Psi|= \lambda^2$, which gives $\yiso_{\ast}\approx \yiso_{a}+(1-\biso_q/\aiso)/(\lambda\Sigma_a^{3/2})$. Thus, $\beta (\riso_{\ast}-\aiso)$ also scales with $\sqrt{e\beta}$ and we know exactly the dependence of the BI potential with respect to these two parameters. Now we can approximate our solution using the fixed value $\zeta=\zeta_a$.

\begin{figure}[t]
	\includegraphics[width=0.47\textwidth,height=0.3\textwidth]{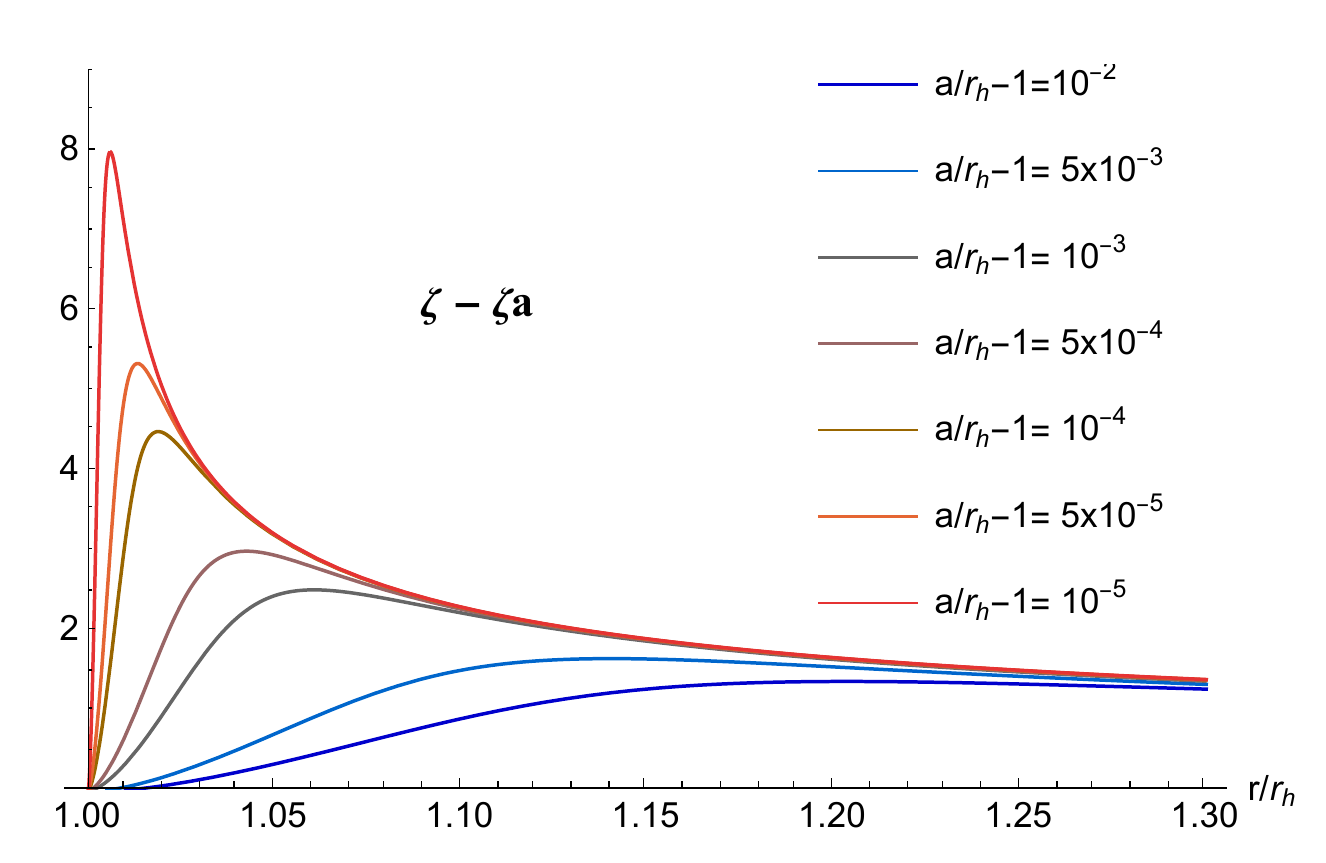}
	\caption{The difference between the function $\zeta$ and $\zeta_a$, showing that $\zeta \geq \zeta_a$ for any value of $\aiso$ and $\riso$.
	}
	\label{fig:zeta_zeta_a}
\end{figure}
Figure~\ref{fig:zeta_zeta_a} shows that $\zeta$ is always greater than $\zeta_a$, hence by fixing the value of $\zeta=\zeta_a$ we are underestimating the value of the potential. This is sufficient for the thought experiment described in the next section since we will establish only a lower limit for the test body energy. Besides, our previous analysis guarantees that the potential scales with $\sqrt{e\beta}$, hence the error due to fixing $\zeta=\zeta_a$ can at most introduce a numerical factor, which is irrelevant for our conclusion. Using \eqref{gradPsiapprox} with $\zeta=\zeta_{a}$, one can directly integrate \eqref{dimlessBI}, which gives
\begin{align}\label{constitBI5}
\phi_{Cop} (\riso,\theta)=
&	
\sqrt{e\beta} \zeta_{a}\Bigg(
\frac{\Gamma\left(\frac14\right)^2 }{4\sqrt{\pi}} \\
&-\frac{ \zeta^2_{a}}{\psi_{Cop}} \ {}_2F_1 \left[\frac14,\frac12,\frac54,- \left(\frac{ \zeta_a}{\psi_{Cop}}\right)^4 \right] 
\Bigg)
\nonumber \  ,
\end{align}
where the argument of the gaussian hypergeometric function is understood as a function of $\riso$ for fixed values of $\theta$. The first term is an integration constant chosen so as to cancel the constant term reminiscent of the asymptotic limit of the  hypergeometric function. The gaussian hypergeometric function has two well-defined limits
\begin{align}
\lim_{x\rightarrow 0}&\quad  {\, {}_2F_1\left[\frac14,\frac12,\frac54,-x^4\right]}=1+\mathcal{O}(x^4)\quad ,\label{Hyperlim0}\\
\lim_{x\rightarrow 0}& \quad \frac1x \, { \, {}_2F_1 \left[\frac14,\frac12,\frac54,-\frac{1}{x^4}\right]}=\frac{\Gamma\left(\frac14\right)^2}{4\sqrt{\pi}}-x+\mathcal{O}(x^5) \quad .\label{Hyperlim1/0}
\end{align}

In the limit $\riso\rightarrow \infty$, the second term of \eqref{constitBI5} tends to $\psi_{Cop}$ plus a constant, which is exactly the first term added to cancel it. In this manner, far away from the charged particle we re-obtain a maxwellian behavior (see~\cite{Falciano2019} for more details). At the particle position where $\psi_{Cop}\rightarrow \infty$ we have
\begin{align}\label{psiCopata}
\phi_{Cop} (\aiso,0)&=  \frac{\Gamma\left(\frac14\right)^2\sqrt{e\beta}}{4\sqrt{\pi}}  \sqrt{\frac{(1-\biso_q/\aiso)}{\Sigma (\aiso)}} \quad .
\end{align}

The first term equals the BI electrostatic potential evaluated at the particle's position for the flat spacetime case (see eq.~(34) of~\cite{Falciano2019}). It is also formally equivalent to the black hole potential \eqref{PhiBH} evaluated at the origin, i.e. 
\begin{equation}\label{defphie}
\phi_{e}(0)=\frac{\Gamma\left(\frac14\right)^2}{4\sqrt{\pi}} \sqrt{e\beta}\quad .
\end{equation}
The second term $\sqrt{(1-\biso_q/\aiso)/\Sigma(\aiso)}=\sqrt{g_{00}(\aiso)}$ accounts for the redshift of the potential due to the spacetime curvature. Thus we can write  
\begin{equation}\label{phiCopphie}
\phi_{Cop}(\aiso,0)=\phi_{e}(0)\sqrt{g_{00}(\aiso)}\quad .
\end{equation}

\section{Lowering a charged test body into EBI-BH}\label{Lowering}

Our thought experiment consists in slowly lower a small spherical object with mass $m$ and charge $e$ as close as possible to the horizon and then drop it into the black hole. We assume this small body to be a test particle in order not to perturb the black hole spacetime, specifically $m/M\ll 1$ and $|e/q|\ll 1$. After the small body is absorbed, the area of the black hole increases and consequently also its entropy. We shall calculate the minimum change in the area of the black hole following~\cite{Christodoulou1970,Carter1968,Misner1974,Bekenstein1999}. The action for the test particle associated with the motion $x^\mu(\tau)$ is
\begin{equation}\label{AcP}
S=\int d\tau\left(mc\sqrt{g_{\mu\nu}\dot{x}^\mu\dot{x}^{\nu}}+\frac{e}{c}\dot{x}_\mu A^{\mu}\right)\ ,
\end{equation}
where a dot means time derivative with respect to proper time $\tau$ and the four-velocity is normalized, i.e. $g_{\alpha\beta}\dot{x}^\alpha\dot{x}^\beta=c^2$. The $A^{\mu}$ is the total electromagnetic potential vector. It has two contributions, one being the background potential due to the black hole $A_0^{(q)}$ and another from the self-potential of the test particle $A_0^{(e)}$, which in curved spacetimes gives a nontrivial contribution to the energy as measured at infinity. 

Given the stationarity of the spacetime, the time-like Killing vector $\xi^\mu$ allow us to define the conserved quantity $\mathcal{E}=c\, p_\mu \xi^\mu$, which can be interpreted as the energy measured at infinity. The contribution of electromagnetic potential to $\mathcal{E}$ is calculated at the particle position, which in the Maxwellian case needs to be regularized by absorbing the divergent coulomb term in the particle's mass. However, since BI is a theory that by construction has a maximum value of electromagnetic fields, the potential is finite everywhere including at the particle's location. The slowly lowering of the particle gives a four-velocity almost stationary, namely $\dot{x}^\mu \approx c(1/\sqrt{g_{00}} \, ,\, \vec{0} )$, while the momentum is defined as usual $p_\mu=\p \mathcal{L}/\p \dot{x}^\mu$, i.e. 
\begin{align}\label{Eptest}
\mathcal{E}=& p_{\mu}\xi^{\mu}c=mc\dot{x}^{\beta}g_{0\beta}+e A_0\left(\aiso,0 \right) \ , \\	
=&mc^2	\sqrt{g_{00}(\aiso)}	+e\phi_q(\aiso)+\frac{e}{2} \Big(\phi_{Cop}(\aiso,0)+\psi_p \left(\aiso\right)\Big)\ . \nonumber
\end{align}
In the above equation, we have  explicitly expressed $A_0$ as a sum of the electromagnetic potential of the black hole $\phi_q(\aiso)$ (see \eqref{PhiBH}) and of the test particle $\phi =\phi_{Cop}+\psi_p$ (see \eqref{eq:phiapproximation}). The caveat is that the latter gains a factor $\frac12$ since only part of it goes to the self-energy while the rest contributes to the background~\cite{Bekenstein1999,Vilenkin1979,Smith_Will1980}. In isotropic coordinates, the horizon is located at $\riso_h=r_h/2-r_s/4$. 
When the particle is close to the horizon, the proper distance of its center of mass to the horizon is~\cite{Bekenstein1999}
\begin{align}
l& \equiv\int_{ \riso_h}^{\aiso}{\dd r \sqrt{g_{\riso \riso}}} 
\approx  \Sigma\left(\riso_h\right)(\aiso-\riso_h)+\mathcal{O}\left(\aiso-\riso_h\right)^2 \quad .
\end{align}

In first order we have $\aiso\Sigma\left(\aiso\right)=a=r_h+\mathcal{O}(l^2)$,
\begin{align*}
\Sigma\left(\aiso\right)&=\frac{4r_h}{2r_h-r_s}\left(1-\frac{l}{r_h}\right)+\mathcal{O}\left(l^2\right)\\
\biso_q/\aiso &=(1+\delta \aiso/\riso_h)^{-2}=1-2\frac{l}{r_h}+\mathcal{O}\left(l^2\right)
\end{align*}
and thus the energy reads
\begin{align*}
\mathcal{E}=&\frac{\left(2r_h-r_s\right)mc^2l}{2r_h^2}+e\phi_q(\aiso)+\frac{e^2r_s}{4r_h^2}
+\frac{e^2}{r_e}\frac{\Gamma\left(\frac14\right)^2}{8\sqrt{\pi}} \sqrt{\frac{l}{2r_h}}\ .
\end{align*}
We obtain a lower bound for the energy by replacing the proper distance $l$ by the proper radius $\mathcal{R}$ of the test body. Since $l\geq \mathcal{R}$ the particle's energy satisfies
\begin{align}\label{BoundE}
\mathcal{E} \geq &
\frac{\left(2r_h-r_s\right)mc^2\mathcal{R}}{2r_h^2}+e\phi_q(\aiso)+\frac{e^2r_s}{4r_h^2}\\
&+\frac{e^2}{4r_h}
\frac{\Gamma\left(\frac14\right)^2}{2\sqrt{\pi}} \sqrt{\frac{r_h\mathcal{R}}{2r_e^2}}+\mathcal{O}(3)
\quad ,
\nonumber
\end{align}
where $\mathcal{O}(3)$ means third order in any combination of $l/M$ or $q/M$. Once the spherical body is absorbed, the black hole's area increases due to a change in its mass $M\rightarrow M+\mathcal{E}/c^2$ and in its charge $q\rightarrow q+e$. The initial area of the black hole is $A=4\pi r_h^2$ and we have two limiting cases to consider: the deep BI regime $(\beta\ll 1)$ and the superficial BI regime $(\beta\gg 1)$.

In the deep BI regime, the electrostatic potential for the black hole (Eq. \eqref{PhiBH}) takes the form
\[
\phi_{q}(\aiso)=\frac{\Gamma\left(\frac{1}{4}\right)^2}{4\sqrt{\pi}}\sqrt{q\beta}+\mathcal{O}\left(\beta\right)\ .
\]
In addition, the metric is similar to SBH but with a rescaled horizon radius $r_h=r_{qs}$ (see \eqref{deepBImetric}). Once the black hole absorbs the test particle its charge changes to $q+e$, hence $\delta A=8\pi r_{qs}\delta r_{qs}$. Direct calculations show 
\begin{align} 
\delta r_{qs}&=\frac{2G \mathcal{E} }{c^4} -\frac{\Gamma\left(\frac{1}{4}\right)^2}{2\sqrt{\pi}}\frac{Ge\sqrt{q\beta}}{c^4}+\mathcal{O}(2) \label{eq:deltahorS}\\
\delta A&=
\frac{16\pi G}{c^4}r_s
\left(\mathcal{E}-
\frac{\Gamma\left(\frac{1}{4}\right)^2}{4\sqrt{\pi}}e\sqrt{q\beta}
\right) +\mathcal{O}(3) \label{eq:dHADeep}\\
&\geq \frac{4\pi G}{c^4}
\Bigg[ 2 mc^2\mathcal{R}  + e^2 \left(1+\frac{\Gamma\left(\frac14\right)^2}{4\sqrt{\pi}}\sqrt{\frac{ 2r_s \mathcal{R}}{ r_e^2 }}\right)\Bigg]
+\mathcal{O}(3) 
\ , \nonumber
\end{align} 
where we have used \eqref{BoundE}. Note that the change in the black hole area depends not only on the particle's parameters but also on the black hole mass.

For the superficial BI regime $(\beta\gg1)$, the black hole electrostatic potential is equivalent to a charge $q$ at the origin, hence at the test particle location, where $\aiso \gg r_\beta$, the BI nonlinearities can be neglected and $\phi_q(\aiso)\approx {q}/{\aiso\Sigma(\aiso)} $. Thus, using the definition of $\psi_p$ and \eqref{psiCopata}, we find
\begin{align}
\mathcal{E}
=&	\frac{mc^2\left(\aiso-\biso_q\right)+eq+e^2r_s/4\aiso\Sigma(\aiso)}{\aiso\Sigma(\aiso)}
\nonumber\\
&+e\sqrt{e\beta}
\frac{\Gamma\left(\frac14\right)^2}{8\sqrt{\pi}} \sqrt{\frac{(1-\biso_q/\aiso)}{\Sigma(\aiso)}}
\quad  .
\end{align}
Assuming the charge of the black hole to be small compared with its mass $q/M\ll 1$, we can approximate $r_h\approx r_s-{r_q^2f(x_h)}/r_s$ and the black hole area by $A\approx 4\pi \left(r_s^2-2r_q^2f(x_s)\right)$. Since $\mathcal{E}/M\ll 1$ and $e/q\ll 1$, and keeping only first order corrections, the change in the area reads
\begin{align} \label{eq:changeHA}
\delta A= \frac{8\pi G}{c^4} &\left(2r_s\mathcal{E} -(2qe+e^2)f(x_s)\right)+\mathcal{O}(3)\nonumber\\
\geq  \frac{4\pi G}{c^4}&
\Bigg[ 2 mc^2 \mathcal{R}
 +2(2eq+e^2)\left(1-f(x_s)\right) \\
& 
-e^2  \left(1-\frac{\Gamma\left(\frac14\right)^2}{4\sqrt{\pi}}\sqrt{\frac{2r_s  \mathcal{R}}{ r_e^2}}\right) \Bigg]+\mathcal{O}(3)
\quad . \nonumber
\end{align} 

The term that multiplies the $(1-f(x_s))$ can be neglected since the function $f(x)$  tends to 1 for large $x$. Indeed, Taylor expanding we have $1-f(x_s)=\frac{1}{20}(\frac{r_\beta}{r_s})^4+\mathcal{O}(x_s^{-8})$, hence we can neglect this term. Note that the last square root term depends on the body's radius $\mathcal{R}$ and charge $e$, and on the black hole mass through $r_s$. The inequality \eqref{eq:changeHA} reproduces the result of~\cite{Falciano2019} for a BI particle falling into a SBH. It is interesting to note that the electromagnetic coupling between the test particle and the black hole is the same in the two different situations. In the deep BI regime, where the nonlinearities dominate, the corresponding expansion of $\phi_{q}(\aiso)$ adequately cancels the contribution coming from the change of the horizon area. On the other hand, in the superficial BI regime, the black hole is close to a RN-BH and in leading order the nonlinear corrections do not contribute to the growth of area.

The black hole entropy is proportional to the black hole area, hence the absorption of the test body also increases the black hole entropy. Using the GSL we can translate the change in the black hole area to a maximum bound to the test body's entropy. However, this procedure is valid only if the increase in the black hole area depends only on the small body parameters. In the two cases analyzed above, the increase in the black hole area also depends on the black hole mass. A straightforward solution is to use the inequalities $r_s\geq \mathcal{R}$. Therefore we can substitute $r_s\rightarrow \mathcal{R}$ in both inequalities above, and the entropy bound for the two cases read
\begin{equation}\label{entropyb2}
S_{BI}\leq 
\frac{2\pi  k_B}{\hbar c}\left\{
\begin{array}{l}
mc^2\mathcal{R} 
+\frac{e^2}{2}\left(1+\sqrt{2}\frac{\mathcal{R}}{\lambda_{\beta}}\right)\quad \mbox{Deep BI}\qquad \\
\\
mc^2\mathcal{R} +\frac{e^2}{2}\left(-1+\sqrt{2}\frac{\mathcal{R}}{\lambda_{\beta}}\right)\ \mbox{Superficial BI}\\
\end{array}
\right.
\end{equation}
where $\lambda_{\beta}\equiv 4\sqrt{\pi} \Gamma\left(\frac14\right)^{-2} r_e\approx 0.54 r_e$. Note that both regimes violate Bekenstein's bounds~\eqref{eq:BekBound} since the extra terms are positive. Both regimes acquire a term proportional to $e^2\mathcal{R}/\lambda_{\beta}$ but physically they are very different. The deep BI has a well defined limit for $\beta\rightarrow 0$ $(\lambda_{\beta}\rightarrow\infty)$ with the BI nonlinearities giving only a small corrections to the entropy. On the other hand, this term becomes very large in the superficial BI regime. Since it is proportional to $\sqrt{\beta}$,  the BI nonlinear term dominates and gives a very large contribution to the entropy bound. In this case, the limit $\beta\rightarrow \infty$ is not well defined in entropy bound. This is expected since the $\beta\rightarrow \infty$ corresponds to the Maxwell limit where the Coulomb potential gives a divergent contribution that is absorbed in the mass of the charge particle. Even though the BI parameter can be made as large as wanted, the bound \eqref{entropyb2} is valid for finite $\beta$. 

It is also worth mentioning the change of sign in the particle's electrostatic self-energy for the deep BI regime, i.e. we get the opposite sign as compared to \eqref{eq:BekBound}. One could naively argue that the deep BI regime ($\beta\ll 1$) makes all charges ineffective and we could not have any charge contribution to the entropy. But this is not correct since the black hole polarization is independent of the NLED. The horizon works as a conducting surface, hence close to the horizon the NLED tends to Maxwell's electrodynamics. In other words, any NLED with the correct Maxwell weak field limit polarize the black hole horizon in the same manner, or equivalently the image charge as seen from outside the black hole always looks like a maxwellian charged particle. The total contribution is a combination of the electromagnetic potential of the black hole, of the test particle and of the imagine particle and the final result clearly depends on the $\beta$ parameter.


\section{Conclusion}\label{conclusion}

The Bekenstein's entropy bound is accepted as a universal inequality relating the physical quantities of an isolated system such as the energy, the angular-momentum, the charge and its size. Additionally, It is assumed to be valid irrespective of the dynamics of the system and conjectured impossible to be improved~\cite{Bekenstein1999}.

We study the same thought experiment by lowering a charge test body close to the horizon and then letting it fall into the black hole. The charged particle satisfies BI electrodynamics in the spacetime of a EBI-BH. Due to consistency, it is essential to have the black hole and the charge test body comply with the same NLED. Indeed, the nonlinearity encoded in the EBI-BH modify the change in the black hole area due to the absorption of the falling test body.

As a side result, we showed (see theorem~\ref{thmNLEDSSBH}) that, for an arbitrary NLED, the displacement vector is curl-free in any static spherically symmetric spacetime and the electrostatic potential can be written in terms of Linet's solution, which is the maxwellian electrostatic potential in Schwarzschild. Using this result, we construct the electrostatic potential of a test BI charged particle in the spacetime of a EBI-BH.

The EBI-BH has two limiting situation: the deep BI regime when $\beta \ll 1$ and the superficial BI regime when $\beta \gg 1$. In the former case, the EBI-BH is very similar to a SBH but with a rescaled mass, i.e. the horizon is located at $r_h=r_{qs}$ (see~\eqref{deepBImetric}), while in the superficial regime, the EBI-BH is close to a RN-BH. In both situation the Bekenstein entropy bound is violated but the coupling with the black hole charge differs. 

The deep BI regime has a small contribution from the BI nonlinearities but the particle's electrostatic self-energy term has the opposite sign as compared to \eqref{eq:BekBound}. In the limit $\beta\rightarrow0$, the BI black holes tends to a Schwarzschild black hole but the polarization of the horizon leaves an imprint by changing the sign of $e^2$. 

The superficial BI regime has a distinct behavior. The particle's electrostatic self-energy remains the same but the BI nonlinearities dominate and give a large positive. In addition, the extra terms contributing to the change in the black hole area combine the particle's parameter with the black hole mass. This is in contrast to the previous analysis of Bekenstein and collaborators where the increase in the black hole area depends only on the particle's parameters. This might suggest that the minimal coupling of matter field with gravity is not sufficient to avoid long-range interaction between the black hole and the matter field. Finally, our analysis shows that, in general, any successful entropy bound must take into account the dynamics of the system.


\section*{Acknowledgments}
The authors would like to thank and acknowledge financial support from the National Scientific and Technological Research Council (CNPq, Brazil), the State Scientific and Innovation Funding Agency of Esp\'\i rito Santo (FAPES, Brazil), and the Brazilian Federal Agency for Support and Evaluation of Graduate Education (CAPES, Brazil). MLP acknowledges support from the State Scientific Funding Agency of Rio de Janeiro (FAPERJ, Brazil).

\appendix

\section{Translating Copson solution in isotropic to standard coordinate}
Copson solution was originally expressed in isotropic coordinates as 
\begin{align}\label{app:linetschw}
\psi_{Cop} (\riso,\theta)&=\frac{e\left(1+r_s/4\aiso\right)^{-2}}{\riso\left(1+r_s/4\riso\right)^2}\left(\mu+\frac{\biso}{\aiso\mu} \right)\ ,
\end{align}
where
\begin{align}
\mu(\riso,\theta)&=\sqrt{\frac{\left(\riso-\biso\right)^2+2\biso\riso\left(1-\cos\theta\right)}{\left(\riso-\aiso\right)^2+2\aiso\riso\left(1-\cos\theta\right)}}\ ,	\  \biso\equiv \frac{r_s^2}{16\aiso}\ .\label{app:mufunct}
\end{align}
The transformation from isotropic to standard coordinate $(\riso \longleftrightarrow r)$ reads
\begin{align}\label{app:std_iso}
&2 \riso= r+\sqrt{r(r-r_s)}-\frac{r_s}{2}\quad , \quad
r=\riso\left(1+r_s/4\riso\right)^2
\end{align}
which implies the following relations
\begin{align}
&r-\frac{r_s}{2}= \riso+\frac{\aiso \biso}{\riso} \quad , \quad \left(r-\frac{r_s}{2}\right)^2-\frac{r_s^2}{8}=\riso^2+\frac{\aiso^2 \biso^2}{\riso^2}\label{app:std_iso2}\\ 
&\aiso+\biso=a-\frac{r_s}{2} \quad , \quad \aiso^2+\biso^2=\left(a-\frac{r_s}{2}\right)^2-\frac{r_s^2}{8}\label{app:std_iso3}
\end{align}

Note that 
\begin{align*}
&\Big(\aiso\mu +\frac{\biso}{\mu}\Big)^2=\frac{\aiso^2\riso^2+\aiso^2\biso^2-2\aiso^2\biso\riso\cos\theta}{\riso^2+\aiso^2-2\aiso\riso \cos\theta }\\
&\hspace{2.cm} +\frac{\biso^2\riso^2+\aiso^2\biso^2-2\biso^2\aiso\riso\cos\theta}{\riso^2+\biso^2-2\biso\riso \cos\theta }+2\aiso\biso\\
&=
\frac{1}{4}\Big((2a- r_s )(2r- r_s)- r_s^2\cos\theta 	\Big)^2
\Big[ \left(2r- r_s \right)^2- r_s^2\\
&+\left(2a-r_s \right)^2-2(2a- r_s)\left(2r- r_s \right)\cos \theta+r_s^2\cos^2\theta 
\Big]^{-1}
\end{align*}
Therefore, straightforward calculation shows that
\begin{align}
\psi_{Cop} (r,\theta)
&=\frac{e}{2ar}
\Big((2a- r_s )(2r- r_s)- r_s^2\cos\theta \Big) \nonumber\\
&\times \Big[ \left(2r- r_s \right)^2+\left(2a-r_s \right)^2-r_s^2\sin^2\theta \nonumber\\
&\qquad -2(2a- r_s)\left(2r- r_s \right)\cos \theta\Big]^{-1/2}\label{app:CopsonSol}
\end{align}

\section{Electrostatic Potential in a Reissner-Nordstr\"om form Copson solution}\label{secappARN}
The differential equation for the electrostatic potential in RN reads
\begin{align}\label{app:eq.psi}
\frac{1}{r^2}\frac{\p }{\p r}\left(r^2\frac{\p \psi}{\p r}\right)+\frac{\hat{L}^2\psi}{\left(r-r_+\right)\left(r-r_-\right)}=-4\pi j^0	\quad ,
\end{align}
where $\hat{L}^2$ is the angular operator
\[
\hat{L}^2\psi=
\frac{1}{\sin\theta}\frac{\p }{\p \theta}\left( \sin \theta \frac{\p \psi}{\p \theta}\right)
+\frac{1}{\sin^2 \theta}\frac{\p^2 \psi}{\p \varphi^2}
\]

For SBH the denominator below $\hat{L}^2$ is $r^2(1-r_s/r)$, hence we want to transform this term into $x(x-C^{te})$. A possible transformation is $x=r-r_-$ such that $(r-r_+)(r-r_-)=x\left[x-(r_+-r_-)\right]$. The other step is to redefine the potential $\psi(r,\theta)=A(r,x)\psi_s(x,\theta)$ to conform the radial operator. Since $\hat{L}^2\psi=A\hat{L}^2\psi_{s}$ the condition is
\begin{align}\label{app:eq:transA}
\frac{1}{r^2}\frac{\p }{\p r}\left(r^2\frac{\p }{\p r}\left(A\psi^{s}\right)\right)=
\frac{A}{x^2}\frac{\p }{\p x}\left(x^2\frac{\p \psi^{s}}{\p x}\right) \quad .
\end{align}

Using $r=x+r_-$ and $\p_r=\p_x$ we find that if $A=(1-r_-/r)=x/r$ \eqref{app:eq:transA} is automatically satisfied. Therefore, the desired transformation is
\begin{align}\label{app:transRNSch}
r=x+r_-\ , \ \psi(r,\theta)=\frac{r-r_-}{r}\, \psi^s(x,\theta) \quad .
\end{align}

Copson solution \eqref{app:CopsonSol} gives the $\psi_{s}(x,\theta)$ part of the electrostatic potential in RN. Note that we have to change $x$ back to $r=x+r_-$ and shift $r_s\rightarrow \sqrt{r_s^2-4r_q^2}=r_s-2r_-$, hence $(2x-r_s) \rightarrow (2r-r_s)$. The final adjustment is to shift the position of the charge due to the modification in the term multiplying $j^0$. The source is a delta function located at $r=a$ but after the change of variable appears a term $x(x+r_-)$, instead of a simple $x^2$. Effectively, this implies that we have to shift $a\rightarrow (a-r_-)$, hence we also have $(2a-r_s) \rightarrow (2a-r_s)$. Implementing all this substitutions, the Copson solution for RN reads
\begin{align}\label{app:CopsonRN}
\psi_{Cop} (r,\theta)
&=\frac{e}{2ar}
\Big( (2a- r_s )(2r- r_s)- (r_s^2-4r_q^2)\cos\theta \Big)\nonumber\\
&\times \Big[ 
\left(2r- r_s \right)^2+\left(2a-r_s \right)^2-(r_s^2-4r_q^2)\sin^2\theta \nonumber\\
&-2(2a- r_s)\left(2r- r_s \right)\cos \theta
\Big]^{-1/2}
\ .
\end{align}
The above solution does not have the correct asymptotic behavior. Indeed,
\[
\lim_{r\rightarrow \infty}\psi_{Cop} (r,\theta)=\frac{e}{r}\left(1-\frac{r_s}{2a}\right)+\mathcal{O}\left(r^{-2}\right) \quad ,
\]
showing that it does not describe a particle of charge $e$ but two charges, one with $e$ and the other with $-e{r_s}/{2a}$. In order to correct the above solution Linet~\cite{Linet1976} showed that it is sufficient to sum a term to cancel this contribution. Thus, the electrostatic potential of a particle of charge $e$ close to a RN-BH reads
\begin{align}\label{app:LinetRN}
\psi(r,\theta)=\psi_{Cop} (r,\theta)+\frac{er_s}{2ar}\quad ,
\end{align}
with $\psi_{Cop} (r,\theta)$ given by \eqref{app:CopsonRN}.

\section{Electrostatic Potential in RN using Isotropic Coordinates}\label{secappIsoCoord}
In standard coordinate system, Copson solution is given by \eqref{app:CopsonRN}. The coordinate transformation between standard $(t,r,\theta,\phi)$ to isotropic $(t,\riso,\theta,\phi)$ is given by
\begin{align*}
&2 \riso= r-\frac{r_s}{2}+\sqrt{r(r-r_s)+r_q^2}\quad , \quad	r=\riso\, \Sigma(\riso)\\
&\aiso+\biso_q=a-\frac{r_s}{2} \quad , \quad \biso_q=\frac{r_s^2-4r_q^2}{16\aiso}\\
\end{align*}
where $\biso_q$ is the generalization of $\biso$ and tends to it in the limit $q\rightarrow 0$. Note that by defining
\[
\mu(\riso)=\sqrt{\frac{\riso^2+\biso_q^2-2\biso_q\riso\cos\theta}{\riso^2+\aiso^2-2\aiso\riso\cos\theta}} \quad ,
\]
the second fraction on the r.h.s. of \eqref{app:CopsonRN} equals $(\aiso\mu(\riso)+\biso_q/\mu(\riso))$. Therefore, comparing with \eqref{app:CopsonRN}, we see that Copson solution for RN using isotropic coordinate reads
\begin{align}\label{app:CopsonRNiso}
\psi_{Cop} (\riso,\theta) &=\frac{e}{\riso}\frac{\Sigma^{-1}(\aiso)}{\Sigma(\riso)}\left(\mu(\riso)+\frac{\biso_q}{\aiso\mu(\riso)}\right)
\end{align}
and the electrostatic solution with the correct asymptotic limit (Linet's solution \eqref{app:LinetRN}) in terms of the isotropic coordinate is
\begin{align}\label{app:LinetRNiso}
\psi(\riso,\theta)=&\frac{e}{\riso}\frac{\Sigma^{-1}(\aiso)}{\mu(\riso)\Sigma(\riso)}\left[\left(\mu(\riso)+\frac{r_s}{4\aiso}\right)^2-\frac{r_q^2}{4\aiso^2}\right] \quad .
\end{align}

\section{Mathematical expansions}\label{app:mathapprox}
Let's analyze the divergent behavior of $\psi_{Cop}$ for the limit approaching the charged particle, namely for $\theta=0$ and $\riso\rightarrow a$. Recall that $\psi =\psi_{Cop}+	\psi_{p}$ with
\begin{align*}
&\psi_p(\riso,\theta)=\frac{er_s}{2}\frac{\left(\aiso+\biso_q+r_s/2\right)^{-1}}{\left(\riso+\aiso\biso_q/\riso+r_s/2\right)}\\
&\psi_{Cop}(\riso,\theta)=\frac{e\aiso\left(\aiso+\biso_q+r_s/2\right)^{-1}}{\left(\riso+\aiso\biso_q/\riso+r_s/2\right)}\left[\mu+\frac{\biso_q}{\aiso\mu}\right]\\	
&\mu=\sqrt{\frac{(\riso-\biso_q)^2+2\riso\biso_q(1-\cos\theta)}{(\riso-\aiso)^2+2\aiso\riso(1-\cos\theta)}} \quad , \quad \biso_q=\frac{r_s^2-4r_q^2}{16\aiso}\\
\end{align*}
Over the particle $\psi_p$ is finite but since $\mu$ diverge $\psi_{Cop}$ is also divergent. Approaching the charged particle $(\riso\rightarrow \aiso, \theta \rightarrow 0)$ we can expand the  relevant quantities as $\riso=\aiso+\delta$ and $\varepsilon=2(1-\cos \theta)$ giving 
\begin{align*}
\mu^2(\riso,\theta)&=\frac{(\aiso-\biso_q)^2+2\delta(\aiso-\biso_q)+\biso_q\varepsilon(\aiso+\delta)}{\delta^2+\aiso(\aiso+\delta)\varepsilon}\\
\frac{1}{\mu^3}\frac{\partial \mu}{\partial \riso}&=
-\frac{\delta}{(\aiso-\biso_q)^2}+\frac{3\delta^2}{(\aiso-\biso_q)^3}+\frac{\aiso(\aiso+\biso_q )}{2(\aiso-\biso_q )^3}\varepsilon+\mathcal{O}\left(\delta^3,\delta\varepsilon\right)\\
\frac{1}{\mu^3}\frac{\partial \mu}{\partial\theta}&=
-\frac{\aiso\sqrt{\varepsilon}}{(\aiso-\biso_q )^2}\left(\aiso-\frac{(\aiso+\biso_q )\delta}{(\aiso-\biso_q )}\right)+\mathcal{O}\left(\delta^2\sqrt{\varepsilon},\varepsilon^{3/2}\right)
\end{align*}

The derivatives of the potential read
\begin{align}
\frac{1}{\psi}\frac{\partial \psi}{\partial {r}}&=	
\frac{-\left(1-\aiso\biso_q/\riso^2\right)}{\left(\riso+\aiso\biso_q/\riso+r_s/2\right)}
+\frac{\left(\aiso^2 \mu^2-\aiso\biso_q \right)}{\left(\aiso \mu+r_s/4\right)^2-r_q^2/4}\frac{1}{\mu}\frac{\partial \mu}{\partial {r}}
\nonumber\\
\frac{1}{\psi}\frac{\partial \psi}{\partial\theta}&=
\frac{\left(\aiso^2 \mu^2-\aiso\biso_q \right)}{\left(\aiso \mu+r_s/4\right)^2-r_q^2/4}\frac{1}{\mu}\frac{\partial \mu}{\partial \theta}
\end{align}

Therefore, in the limit approaching the particle we have $\psi \approx \psi_{Cop}$ and 
\begin{align}
\frac{\psi_{Cop}}{\mu} &\approx \frac{e\, \Sigma^{-1}(\aiso)}{\riso\, \Sigma(\riso)}\label{app:psiapprox}
\quad , \quad |\boldsymbol{\nabla} \mu| \approx  \frac{ \mu^2}{\Sigma(\riso)}\frac{(\aiso-\biso_q)}{(\riso-\biso_q)^{2}}\nonumber\\
\Sigma(\riso)=&\left(1+\frac{r_s}{4\riso}\right)^2-\frac{r_q^2}{4\riso^2}\\
\boldsymbol{\nabla}  \psi_{Cop} &\approx \frac{\psi_{Cop}}{\mu} \boldsymbol{\nabla}  \mu  \quad \Rightarrow \quad |\boldsymbol{\nabla} \psi_{Cop}| \approx  \beta \, \left(\frac{\psi_{Cop}}{\zeta}\right)^2\nonumber\\
\mbox{with } \quad&
\zeta^2\equiv \frac{\beta e(1-\biso_q/\aiso)}{\Sigma\left(\aiso\right)}
\end{align}

\section{Potential Close to the horizon}\label{secappPotCloseHorizon}

The metric in isotropic coordinates reads
\begin{align}
	\dd s^2=&\left(1-\frac{r_s^2-4r_q^2f(x_h)}{16\riso^2}\right)^2\frac{c^2\dd t^2}{\Sigma^2}-\Sigma^2\left[\dd \riso^2+\riso^2\dd \Omega^2\right]\ , \nonumber\\
	\mbox{with}& \quad \Sigma(\riso)=\left(1+\frac{r_s}{4\riso}\right)^2-\frac{r_q^2f(x_h)}{4\riso^2} \quad . \label{appSigmaRN}
\end{align}
and the potential is
\begin{align*}
&\psi(\riso,\theta)=\frac{e}{\riso}\frac{\Sigma^{-1}(\aiso)}{\mu(\riso,\theta) \Sigma(\riso)}\left[\left(\mu(\riso,\theta)+\frac{r_s}{4\aiso}\right)^2-\frac{r_q^2f(x_h)}{4\aiso^2}\right] \nonumber\\
&\mu(\riso,\theta)=\sqrt{\frac{\riso^2+\biso_q^2-2\biso_q\riso\cos\theta}{\riso^2+\aiso^2-2\aiso\riso\cos\theta}} \ , \ \biso_q=\frac{r_s^2-4r_q^2f(x_h)}{16\aiso}
\end{align*}
where $\aiso$ is the position of the charged particle. Note that when the particle gets close to the horizon, namely $\aiso=(1+\delta)\riso_h$ with $0<\delta \ll 1$, we have $\biso_{q}=\riso_h^2/\aiso=\riso_h(1-\delta)+\mathcal{O}(\delta^2)$ and 
\begin{align}
&\mu(\riso,\theta)=1+2\delta \frac{\riso_h(\riso-\riso_h-\riso(1-\cos\theta))}{(\riso-\riso_h)^2+2\riso_h\riso(1-\cos\theta)}+\mathcal{O}\left(\delta ^2\right) \ .
\end{align}


\end{document}